\documentclass[
    ,final            
  ]
  {aipproc}

\layoutstyle{6x9}

\usepackage{graphicx}

\begin{document}
\begin{flushright} UK/08-04 \end{flushright}
\title{Challenges of Lattice Calculation of Scalar Mesons}



\author{Keh-Fei Liu}{
  address={Dept. of Physics and Astronomy, University of Kentucky,
  Lexington, KY 40506 USA}
}

\begin{abstract}
 I review a proposed pattern of the light scalar mesons
 with $q\bar{q}$ mesons and glueball above 1 GeV  and tetraquark mesoniums
 below 1 GeV. Several challenges and caveats of calculating
 these light scalar mesons with dynamical fermions are discussed.
\end{abstract}

\maketitle


\section{Introduction}

\begin{figure}[t]
 \resizebox{.7\textwidth}{!}
  {\includegraphics{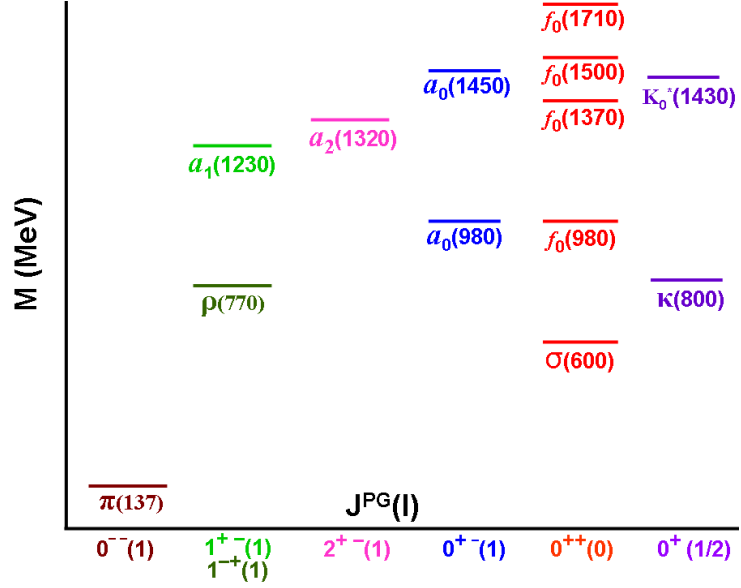}}
  \caption{Spectrum of scalar mesons together with
 $\pi$, $\rho, a_1$ and $a_2$.}
 \label{fig:scalar}
\end{figure}

The pseudoscalar, vector, axial, and tensor mesons with light
quarks (i.e. $u,d$ and $s$) are reasonably well known in terms of
their $SU(3)$ classification and quark content. The scalar meson
sector, on the other hand, is much less understood in this regard.
There are 19 experimental states below 1.8 GeV which are more than
twice the usual $q\bar{q}$ nonet in other sectors. We show in
Fig.~\ref{fig:scalar} the experimentally known scalars including
$\sigma(600), \kappa(800)$, and $f_0(1710)$ which are better
established experimentally nowadays~\cite{pdt06,ait01}. The recent
theoretical advance~\cite{ccl06a} in identifying $\sigma(600)$ as
a $\pi\pi$ resonance by solving the Roy equation has settled the
question about the existence of $\sigma(600)$. Nevertheless, there
are still a number of puzzling features regarding the ordering of
$a_0(1450)$ and $K_0^*(1430)$ with respect to their counterparts
in the axial-vector and tensor sectors, the narrowness of
$a_0(980)$ and $f_0(980)$ in contrast to the broadness of
$\sigma(600)$ and $\kappa(800)$, etc~\cite{liu07}. We shall first
review a emerging pattern of the scalar mesons below 1.8 GeV based
on quenched lattice calculation and phenomenology and then discuss
the challenges and caveats of full QCD calculation of these scalar
mesons on the lattice.

\section{Pattern of Light Scalar Mesons}

The unsettling features regarding the nature of $a_0(1450)$ and
$K_0^*(1430)$ are tentatively resolved in a recent quenched
lattice calculation~\cite{mac06} with overlap fermions for a range
of pion masses with the lowest one at 180 MeV. When the quenched
ghost states, which correspond to $\pi \eta$ and $\pi\eta'$
scattering states in the dynamical fermion case are removed, it is
found that $a_0$ is fairly independent of the quark mass. In other
words, below the strange quark mass, $a_0$ is very flat and
approaches $a_0(1450)$ in the chiral limit. This suggests that
$SU(3)$ is a much better symmetry in the scalar meson sector than
the other meson sectors and that both $a_0(1450)$ and
$K_0^{*}(1430)$ are $q\bar{q}$ states. Furthermore, $f_0(1500)$,
by virtue of the fact that it is close by, should be a fairly pure
$SU(3)$ octet state, i.e.
$f_{octet}=(u\bar{u}+d\bar{d}-2s\bar{s})/\sqrt{6}$.

Based on the lattice findings, a mixing scheme for the isoscalar
$f_0(1370), f_0(1500)$ and $f_0(1710)$ -- a glueball candidate,
with slight $SU(3)$ breaking was developed and successfully fit to
the decays of pseudoscalar meson pairs as well as various decays
from $J/\Psi$~\cite{ccl06b}. Some of the robust and conspicuous
features of this mixing scheme are the following:

\begin{itemize}
\item
       $f_0(1500)$ is indeed a fairly pure octet with very little mixing
with the flavor singlet and the glueball. $f_0(1710)$ and
$f_0(1370)$ are dominated by the glueball and the $q\bar{q}$
singlet respectively, with $\sim 10\%$ mixing between the two.
This is consistent with the experimental result $\Gamma(J/\Psi
\rightarrow \gamma f_0(1710)) \sim 5 \Gamma(J/\Psi \rightarrow
\gamma f_0(1500))$~\cite{ait01} which favors $f_0(1710)$ to have a
larger glueball content.

\item
      The ratio $\Gamma(f_0(1500)\rightarrow  K\overline{K})/\Gamma(f_0(1500)
\rightarrow  \pi\pi) = 0.246\pm0.026$ is one of the best
experimentally determined decay ratios for these
mesons~\cite{pdt06}. If $f_0(1500)$ is a glueball (i.e. a flavor
singlet) or $s\bar{s}$, the ratio will be 0.84 or larger then
unity. Either one is much larger than the experimental result. On
the other hand, if $f_0(1500)$ is $f_{octet}$, then the ratio is
$0.21$ which is very close to the experimental value. This further
demonstrates that $f_0(1500)$ is mainly an octet and its
experimental decay ratio can be well described with a small
$SU(3)$ breaking~\cite{ccl06a}.

\item Because the $n\bar n$ content is more copious than the
$s\bar s$ in $f_0(1710)$ in this mixing scheme, the prediction of
$\Gamma(J/\psi\to \omega f_0(1710))/\Gamma(J/\psi\to \phi
f_0(1710))=4.1$ is naturally large and consistent with the
observed value of $6.6\pm 2.7$. This ratio is not easy to
accommodate in a picture where the $f_0(1710)$ is dominated by
$s\bar{s}$. One may have to rely on a doubly OZI suppressed
process to dominate over the singly OZI suppressed process to
explain it~\cite{ac95} .

\end{itemize}

\begin{figure} [th]
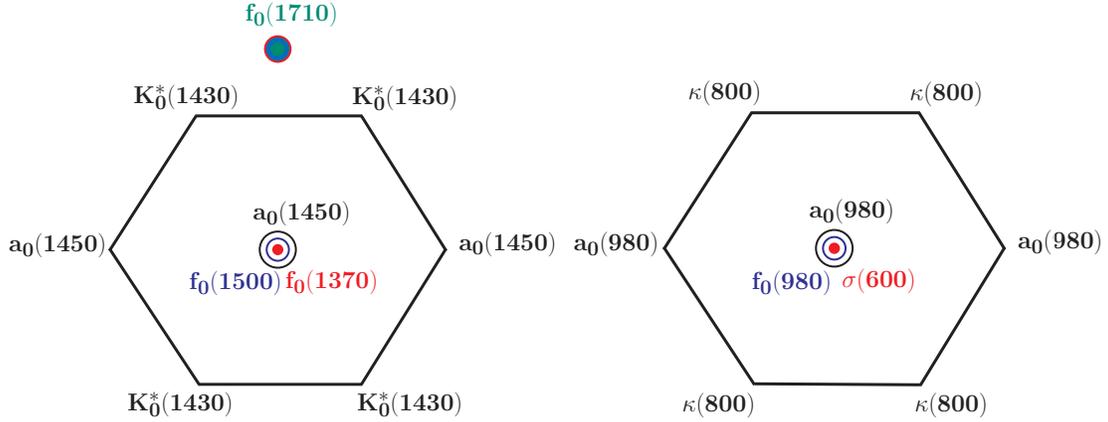

\vspace*{-2.0in}
    \includegraphics[width=7.0cm]{scalar2a.epsi}
\vspace{1.0in}
\hspace*{0.3cm}
       \includegraphics[width=7.0cm]{scalar1a.epsi}
  \caption{Pattern of light scalar mesons -- a tetraquark mesonium nonet below 1 GeV,
an almost pure $SU(3)$ $q\bar{q}$ nonet and a nearly pure glueball
above 1 GeV.
  }
\end{figure}

The mesons below 1 GeV were suggested to be tetraquark
mesoniums~\footnote{These are two-quark and two-antiquark mesons
which have been referred to as four-quark states, meson
moleculars, mesoniums, and tetraquark states. We shall call them
tetraquark mesoniums so as to avoid implication on the nature of
possible spatial and color clustering.} from the MIT bag
model~\cite{jaf77} and potential model~\cite{lw81,wi82} studies. A
recent lattice calculation~\cite{mac06} with the overlap fermion
on $12^3 \times 28$ and $16^3 \times 28$ quenched lattices with
the two-quark-two-antiquark interpolation field
$\overline{\Psi}\gamma_5\Psi\overline{\Psi}\gamma_5\Psi$ has
confirmed the existence of such low-lying scalar tetraquark
mesonium at $\sim 550$ MeV. This strongly suggests that it is the
$\sigma(600)$.

Combining the lattice calculations of $a_0(1450)$, $K_0^*(1430)$
and $\sigma(600)$ and the mixing study of $f_0(1370), f_0(1500)$
and $f_0(1710)$, a classification of the scalar mesons below 1.8
GeV was proposed~\cite{liu07}. Those below 1 GeV, i.e.
$\sigma(600), a_0(980), f_0(980)$ and $\kappa(800)$ form a nonet
of tetraquak mesoniums; those above 1 GeV, i.e. $a_0(1450),
K_0^*(1430)$ and $f_0(1500)$ form a fairly sure $SU(3)$ octet; and
$f_0(1370)$ and $f_0(1710)$ are good $SU(3)$ singlet and glueball
respectively, with $\sim 10\%$ mixture between the two.

We should stress that this is not finalized. It should be
scrutinized in future experiments, such as high statistics
$J/\Psi$, $D$, $B$ decays and $p\bar{p}$ annihilations.
Furthermore, most of the the lattice results which led to the
above proposed pattern were based on quenched calculations. There
are loose ends that need to be tightened, come dynamical fermion
calculations. In the following, we shall enumerate a number of
challenges and the associated caveats in calculations with
dynamical fermion configurations.

\section{Challenges and Caveats of Future Lattice Calculations
with Dynamical Fermions}

    In the quenched lattice calculation of $a_0$ with light
quarks which correspond to $m_{\pi} < 500$ MeV, the quenched $\pi
\eta$ ghost states are lower than the $a_0(1450)$ and, thus,
dominate the long time behavior in the scalar correlator with a
non-unitary negative tail. This has to be
removed~\cite{mac06,bde02,pdi04} before the physical $a_0(1450)$
is revealed. These ghost states turn into physical two meson
scattering states in a full QCD calculation with the same valence
and sea quark masses~\footnote{Otherwise, it is considered to be a
partially quenched calculation.}. This causes some difficulty in
the fitting of scalar meson correlators and has been mentioned by
S. Prelovsek~\cite{pre08} in this workshop. In the following, we
shall point out several caveats and challenges facing the scalar
meson calculation with light dynamical fermions.

\subsection{$a_0(1450)$ and $K_0^*(1430)$}

   There are several $N_f=2$ dynamical fermion calculations of
$a_0$ with the $\overline{\Psi}\Psi$ interpolation field
~\cite{pdi04,sca04,mm06,fri07,hi08}. Save for Ref.~\cite{pdi04}
which, upon removing the partially quenched ghost $\pi \eta_2$
state, found $a_0$ to be at $1.51(19)$ GeV, the
others~\cite{mm06,fri07,hi08} found the lowest states at the
chiral limit to be $\sim 1$ GeV, suggesting that $a_0(980)$ is the
$q\bar{q}$ state. As pointed out in Ref.~\cite{liu07}, this is
most likely an untenable interpretation.  If $a_0(980)$ is indeed
a $q\bar{q}$ state, or has a sizable coupling to the
$\overline{\Psi}\Psi$ interpolation field, then replacing the
$u/d$ quark in the $a_0$ interpolation field with $s$ will place
the corresponding $s\bar{u}$ at $\sim 1100$ MeV. This is far (i.e.
300 MeV) away from each of the two experimental states
$K_0^*(1430)$ and $\kappa(800)$ (see Fig. 1). The likely
resolution, we think, is that the state found at $\sim 1$ GeV is
the $\pi\eta_2$ scattering state. Since $\eta_2$, the
$\eta(\eta')$ in the two-flavor case is predicted to be $\sim
\sqrt{2/3}m_{\eta'}= 782$ MeV in the large $N_c$ analysis with
$U(1)$ anomaly, the weakly interacting $\pi\eta_2$ will be near
the state seen at $\sim 1$ GeV. In other words, this $\pi\eta_2$
scattering state is the dynamical fermion realization of the
corresponding ghost state in the quenched approximation. Parallel
to the lesson learned in lattice calculations of pentaquark
baryons~\cite{liu06}, one has to include the multi-hadron states
in addition to the physical resonances when fitting the two-point
correlators for the excited spectrum. In the case of $a_0$ in the
realistic $N_f=2+1$ case, one needs to include $\pi \eta,
\pi\eta'$, in addition to the physical $a_0(980)$, and
$a_0(1450)$. This can be achieved with the sequential empirical
Bayes method for curve-fitting~\cite{cdd04} or the variational
approach. Furthermore, one needs to distinguish the two-particle
scattering states from the one-particle resonances. One way to
distinguish a two-particle scattering state from a one-particle
state is to examine the 3-volume dependence of the fitted spectral
weight~\cite{mcd05,mla04,mac06}. Another way is to impose a
`hydrid boundary condition' on the quark propagators~\cite{idi05}.
No attempt has been made to identify the scattering $\pi \eta
(\eta')$ states so far. This has to be carried out before one can
reasonably reveal the quark content of $a_0(980)$ and $a_0(1450)$.

\subsection{$f_0(980), f_0(1370), f_0(1500)$ and $f_0(1710)$}

   In addition to the complication of two-meson scattering
states (in this case $\pi\pi, K\overline{K}, \eta\eta,
\eta\eta'$), one needs to calculate the correlators with
disconnected insertions (D.I.) in addition to the connected
insertions (C.I.) as in the $a_0$ case. This is to reflect the
fact that these isoscalar mesons have annihilation channels. The
usual approach of adopting the noise~\cite{dl94} to estimate the
quark loops in the disconnected insertion makes the calculation
much more expensive than the connected insertion one. One caveat
with the noise estimator is that the signal falls exponentially in
the meson correlator; whereas, the variance of the noise estimator
approaches a constant at certain time separation~\cite{dl94}. If
one were to fit the time window where the variance of the noise
levels off, the shoulder effect of the correlator could result in
an unphysically light effective mass. In view of this, the very
light mass from the D.I. part of the correlator in the $f_0$
calculations~\cite{sca04,hmm06} should be subjected to the
examination as to whether it is the $\pi\pi$ scattering state or
due to the shoulder effect.

\subsection{Glueball}

    The continuum and large volume limits of the quenched
calculation places the scalar glueball at 1710(50)(80)
MeV~\cite{cad06}. This is very close to the viable experimental
glueball candidate $f_0(1710)$. To verify this in the full QCD
calculation is, however, non-trivial. Whatever interpolation one
adopts, one has to disentangle the glueball from all the
lower-lying $f_0$ states and the $\pi\pi, K\overline{K}, \eta\eta$
and $\eta\eta'$ two-meson states.

\subsection{Tetraquark Mesoniums}

   If the nonet below 1 GeV in Fig. 1 are indeed dominated by
the $q^2\bar{q}^2$ tetraquark mesonoums, one can access them
through the
$\overline{\Psi}\gamma_5\Psi\overline{\Psi}\gamma_5\Psi$ operator
or other four-quark operators with the same quantum number. In the
case of $a_0(980)$ and $f_0(980)$, the two-meson threshold, i.e.
$K\overline{K}$ is close by. One may need a good variational
method in order to disentangle them. By virtue of the fact that
$a_0(980)$ and $f_0(980)$ are nearly degenerate, the D.I. should
be small compared to the C.I. It should be confirmed in full QCD
calculation.

\subsection{$q\bar{q}$ meson vs $q^2\bar{q}^2$ tetraquark
mesonium}

   The notion of $q\bar{q}$ or $q^2\bar{q}^2$ meson is primarily a
quark model concept of the valence quark content. How does one
distinguish them in lattice QCD with interpolation fields? So far,
neither $a_0(980)$ nor $\sigma(600)$ is coupled to the
$\overline{\Psi}\Psi$ interpolation field in the quenched
approximation with discernable signal~\cite{mac06}. If this is not
true in full QCD calculation with light dynamical fermions, this
will complicate matters substantially. One will need both
$q\bar{q}$ and $q^2\bar{q}^2$ types of operators with a large
basis in the variational calculation to identify states and;
moreover, to distinguish the one-particle states from the
multi-meson scattering states.

\section{Conclusion}

    We summarized the pattern of light scalar mesons emerged from
quenched lattice calculations and the study of mixing and decays
of $f_0(1370), f_0(1500)$ and $f_0(1710)$. We have discussed the
subtlety and challenges of calculating them in full QCD with light
dynamical fermions. In particular, if they couple strongly to both
$q\bar{q}$ and $q^2\bar{q}^2$ types of interpolation operators,
the interpretation of the underline pattern will be considerably
more complex. We hope that Nature is only subtle but not
malicious.


\begin{theacknowledgments}
  It is a pleasure for the author to thank G. Rupp for his invitation to attend the Scadron70 workshop
  on scalar mesons and his hospitality. He also acknowledges inspiring discussions with D.V. Bugg,
  H. Leutwyler, S. Prelovsek, J. Rosner, and M.D. Scadron.
\end{theacknowledgments}



\bibliographystyle{aipproc}   

\bibliography{sample}

\IfFileExists{\jobname.bbl}{}
 {\typeout{}
  \typeout{******************************************}
  \typeout{** Please run "bibtex \jobname" to optain}
  \typeout{** the bibliography and then re-run LaTeX}
  \typeout{** twice to fix the references!}
  \typeout{******************************************}
  \typeout{}
 }


\end{document}